\documentclass[prb,twocolumn,amscd,amsmath,amssymb,verbatim]{revtex4}

\usepackage{graphicx}
\usepackage[dvips]{hyperref}
\usepackage[TS1,OT1,T1]{fontenc}

\begin{document}
\title{Unusual signatures of the ferromagnetic transition in the heavy Fermion compound UMn$_2$Al$_{20}$}

\author{C. H. Wang$^{1,2,3}$, J. M. Lawrence$^1$, E. D. Bauer$^2$, K. Kothapalli$^4$, J. S. Gardner$^{5,6}$, F. Ronning$^2$, K. Gofryk$^2$,
J. D. Thompson$^2$, H. Nakotte$^4$, F. Trouw$^2$}
\affiliation{$^1$University of California, Irvine, California 92697\\
  $^2$Los Alamos National Laboratory, Los Alamos, NM 87545\\
  $^3$Neutron Scattering Science Division, Oak Ridge National Laboratory, Oak Ridge, TN, 37831\\
  $^4$Department of Physics, New Mexico State University, Las Cruces, 88003\\
  $^5$NCNR, National Institute of Standards and Technology, Gaithersburg, MD 20899-6102\\
  $^6$Indiana University, Bloomington, Indiana 47408
  }

\date{\today}

\begin{abstract}

Magnetic susceptibility results for single crystals of the new cubic
compounds UT$_2$Al$_{20}$ (T=Mn, V, and Mo) are reported.
Magnetization, specific heat, resistivity, and neutron diffraction
results for a single crystal and neutron diffraction and inelastic
spectra for a powder sample are reported for UMn$_2$Al$_{20}$. For T
= V and Mo, temperature independent Pauli paramagnetism is observed.
For UMn$_2$Al$_{20}$, a ferromagnetic transition is observed in the
magnetic susceptibility at $T_c$ = 20 K. The specific heat anomaly
at $T_c$ is very weak while no anomaly in the resistivity is seen at
$T_c$. We discuss two possible origins for this behavior of
UMn$_2$Al$_{20}$: moderately small moment itinerant ferromagnetism,
or induced local moment ferromagnetism.

\end{abstract}

\vskip 15 pt

\pacs{71.27.+a, 75.30.Cr, 75.20.Hr}

\maketitle

\emph{Introduction}  UMn$_2$Al$_{20}$, UV$_2$Al$_{20}$ and
UMo$_2$Al$_{20}$ are members of a new family of lanthanide and
actinide compounds RT$_2$M$_{20}$(R=Ce, Yb, Gd,U; T=transition
metal; M=Zn and
Al)\cite{LnT2Al20,UCr2Al20,UV2Al20,Canfield1,Bauer1,Bauer2,UIr2Zn20}.
These compounds crystallize in the CeCr$_2$Al$_{20}$ type cubic
structure (F d -3 m) and display interesting features such as heavy
fermion or intermediate valence
behavior\cite{Canfield1,Bauer1,Bauer2,UIr2Zn20}. In this structure,
every $f$-atom is surrounded by 16 zinc atoms in a nearly spherical
array of cubic site symmetry, which leads to small crystal field
splittings. Because the R-atom content is less than 5$\%$ of the
total number of atoms, and the shortest $f/f$ spacing is $\sim$ 6
{\AA}, these compounds are valuable for studies close to the
impurity limit but in ordered systems.

We have recently reported on the behavior of the heavy Fermion
paramagnets UCo$_2$Zn$_{20}$\cite{Bauer1,Wang} and
URu$_2$Zn$_{20}$\cite{Wang} as well as of
UIr$_2$Zn$_{20}$\cite{Bauer2}, which exhibits weak itinerant
ferromagnetism. In this report we present the magnetic
susceptibility of UMn$_2$Al$_{20}$, UV$_2$Al$_{20}$ and
UMo$_2$Al$_{20}$, and the specific heat, resistivity, magnetization,
and neutron scattering spectra for UMn$_2$Al$_{20}$. For
UV$_2$Al$_{20}$ and UMo$_2$Al$_{20}$, Pauli paramagnetism is
observed.  For UMn$_2$Al$_{20}$, the magnetic susceptibility shows a
ferromagnetic phase transition at 20 K where the anomaly in the
specific heat is weak and no anomaly is observed in the resistivity.
The neutron diffraction profiles of both polycrystal and single
crystal samples show no obvious extra contribution from the
ferromagnetism below the transition temperature. The inelastic
neutron scattering spectra of a polycrystal sample exhibit no
obvious magnetic excitations in the energy transfer range of 5 to 50
meV. We discuss two possibilities to explain the magnetic behavior
in UMn$_2$Al$_{20}$: heavy Fermion ferromagnetism of itinerant 5$f$
electrons or induced ferromagnetism arising from a low energy
singlet-triplet crystal field excitation of localized 5$f$
electrons.

\emph{Experiment} Single crystals were grown in Al flux with an
elemental starting ratio U:T:Al=1:2:50. Elemental purities were 99.9$\%$ for the (depleted)
U, 99.99$\%$ for the Mn and 99.9999$\%$ for the Al. The crucible was sealed under
vacuum in a quartz tube and was heated to 1050$^0$C quickly in order
to avoid the reaction between Al and the quartz tube. After holding
at 1050$^0$C for 4 h, it was cooled at a rate 5$^0$C/h to 700$^0$C.
At this point the excess Al flux was removed by using a centrifuge.
The magnetization was measured in a commercial superconducting
quantum interference device (SQUID) magnetometer. The specific heat
measurements were performed in a commercial physical properties
measurement system (PPMS). The electrical resistivity was also
measured in the PPMS using the four wire method. The powder neutron
diffraction experiment was performed on the high resolution
diffractometer (BT-1) at the NIST Center for Neutron Research
(NCNR); the sample was a powder
ground from single crystals. The single crystal neutron diffraction experiment was
performed on the single crystal diffractometer (SCD) at the Lujan
Center, LANSCE, at Los Alamos National Laboratory. The inelastic
neutron scattering experiments were performed on a 35 gram powder
sample using the high resolution chopper spectrometer (Pharos) at
the Lujan Center.

\begin{table*}[htp]
\caption{\label{tab:table} Structural parameters of UMn$_2$Al$_{20}$
at room temperature from SCD and at 100 K from BT-1. Error in the
last digit are in the parentheses.}
\begin{ruledtabular}
\begin{tabular}[b]{ccccccc}

space group & $Fd\overline{3}m$& no. 227 & a$^{SCD}$=14.326(6)${\AA}$ &[a$^{BT1}$=14.3190(2)${\AA}$] & $\chi^2_{SCD}$=1.984&$\chi^2_{BT1}$=7.053\\

Atoms & Position & x$^{SCD}$[x$^{BT1}$]  & y$^{SCD}$[y$^{BT1}$]  & z$^{SCD}$[z$^{BT1}$]  & occupancy & Uiso$^{SCD}$($\times$10$^2$)\\
\hline
U   & 8a & 1/8 & 1/8 & 1/8 & 1 & 2.34 \\
Mn  & 16d & 1/2 & 1/2 & 1/2 & 1 & 3.15 \\
Al1 & 16c & 0 & 0 & 0 & 1 & 2.98\\
Al2 & 48f & 0.4893(4)[0.4913(4)] & 1/8 & 1/8 & 1 & 3.56\\
Al3 & 96g & 0.0589(4)[0.0590(1)] & 0.0589(4)[0.0590(1)] & 0.3258(4)[0.3259(1)] &1 & 2.92\\
 &   & $R(F^2)^{SCD}$=4.76$\%$ & $R_w(F^2)^{SCD}$=16.88$\%$ &
 $R_{p}^{BT1}$=14.29$\%$ & $R_{wp}^{BT1}$=17.38$\%$ &

\end{tabular}
\vspace*{-2mm}
\end{ruledtabular}
\end{table*}

\emph{Results and Discussion} The samples were determined to be
single phase within the resolution of the x-Ray, neutron powder, and
neutron single crystal diffraction experiments. Refinements of the
single crystal and powder sample diffraction patterns imply full
occupancy of the atom sites. The results of the refinement are shown
in table I. The magnetic susceptibility $\chi(T)$ of
UMn$_2$Al$_{20}$ is shown in the inset of Fig. 1(a). A dramatic
enhancement at low temperature is observed, indicating a
ferromagnetic transition at $T_c$ $\approx$ 20 K. Due to the small
coercive field(9 Oe, see below), the zero field cooling and field
cooling data is almost the same when the measured field is 0.1 T.
Due to the relatively small magnetic moment of the uranium (0.89
$\mu_B$, see below) and the small fraction of uranium atoms in the
unit cell (less than 5$\%$), the neutron diffraction results on both
the powder and the single crystal samples did not display any
obvious extra intensity at temperatures below the transition
temperature that would correspond to ferromagnetic ordering.

\begin{figure}[t]
\centering
\includegraphics[width=0.4\textwidth]{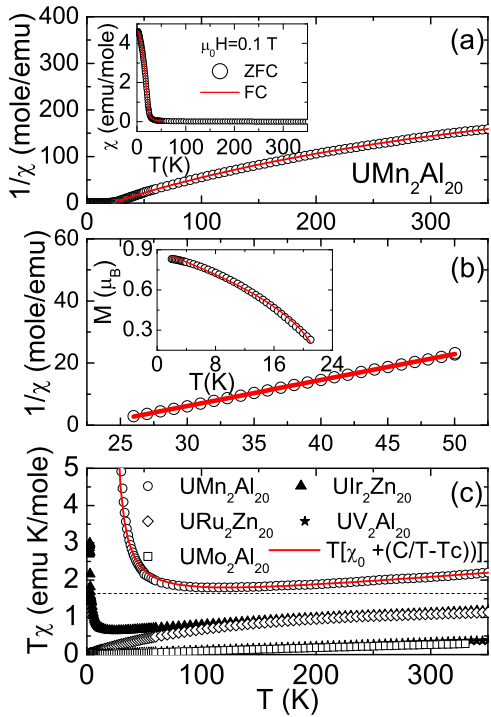}
\caption{\label{fig:1} (a) Inverse magnetic susceptibility
1/$\chi(T)$ for UMn$_2$Al$_{20}$. The line represents the modified
Curie-Weiss fit $\chi(T)= C/(T-\theta)+\chi_0$ with parameters given
in the text. Inset: $\chi(T)$. Open circle is zero-field cooling
curve and solid line is the field-cooling curve. ( b) Mean field
fits for $1/\chi$ in temperature range of 25 K to 50 K (main panel)
and for $M(T)$ below the ferromagnetic transition temperature
(inset) with parameters given in the text. (c) $T\chi$ vs. T for
UMn$_2$Al$_{20}$, UV$_2$Al$_{20}$, UMo$_2$Al$_{20}$,
URu$_2$Zn$_{20}$ (data from Ref.\cite{Wang} ) and UIr$_2$Zn$_{20}$
(data from Ref.\cite{Bauer2} ). The dashed line is the Curie
constant for 5$f^3$ free ion.} \vspace*{-3.5mm}
\end{figure}

The inverse magnetic susceptibility 1/$\chi$(T) of UMn$_2$Al$_{20}$
is shown in Fig. 1(a). A fit of the inverse susceptibility to the
formula $\chi(T)= C/(T-\theta)+\chi_0$ for temperatures above $T_c$
gives  $\theta$=21.2 K, $\chi_0$=0.00253 emu/mole, and C=1.23 emu
K/mole. This value of Curie constant, which is reduced relative the
free ion $f^2$ or $f^3$ Hund's Rule value 1.6 emu K/mole, is typical
of itinerant uranium compounds. The inverse susceptibility
1/$\chi(T)$ in the low temperature range (25 K-50 K) is shown in
Fig. 1(b). The fit to the form $\chi(T)=C/(T-T_c)^{\gamma}$ in this
temperature range yields $T_c$=22.7 K, C=1.17 emu K/mole, and
$\gamma$ = 0.99, which is essentially the mean field critical
exponent ($\gamma$=1). The magnetization $M(T)$ for $T < T_c$ is
shown in the inset of Fig. 1(b). The fit to the formula
$M=M_0(\frac{T_c-T}{T_c})^{\beta}$ yields $M_0$=0.89 $\mu_B$,
$T_c$=22.1 K and $\beta$= 0.49 which is again the mean field theory
exponent ($\beta$=0.5).

The effective moment $T\chi$ versus temperature is compared for
several compounds UT$_2$M$_{20}$(T=Mn, Ir, Ru, V and Mo; M=Al and
Zn) in Fig. 1(c). The dashed line represents the Curie constant for
the Hund's Rule coupled 5$f^2$ or 5$f^3$ free ion. The linear
behavior of $T\chi$ for UV$_2$Al$_{20}$ and UMo$_2$Al$_{20}$
indicates that the magnetic susceptibilities for these two compounds
are essentially temperature independent; the values 0.0011 emu/mole
for UV$_2$Al$_{20}$ and 0.00087 emu/mole for UMo$_2$Al$_{20}$ are
typical of uranium based Pauli paramagnets. Both the heavy Fermion
compound URu$_2$Zn$_{20}$ and the weak itinerant ferromagnet
UIr$_2$Zn$_{20}$\cite{Bauer2} exhibit similar behavior at high
temperature with $T\chi$ of order 1.2 emu K/mole, similar to the
Curie constant observed for UMn$_2$Al$_{20}$ in Figs. 1(a) and (b).
For both UIr$_2$Zn$_{20}$ and UMn$_2$Al$_{20}$, the upturn in
$T\chi$ at low temperatures corresponds to the onset of
ferromagnetic fluctuations, which occur already for $T > T_c$. Given
that the formula $\chi(T)= C/(T-\theta)+\chi_0$ fits the data for
UMn$_2$Al$_{20}$ above $T_c$, and that the Curie constant in this
fit is smaller than the  free ion value, the fact that $T\chi$ is
larger than the free ion value for $T >$ 100 K clearly arises from
the presence of the large constant term  $\chi_0$ $\sim$ 0.0025
emu/mole. A possible explanation for this rather large constant
contribution is that the susceptibility of the manganese atoms is
enhanced. T-independent  susceptibilities  of this order of
magnitude occur, for example, for Mn atoms in alloys of the enhanced
Pauli paramagnet YMn$_2$\cite{Hauser}.

\begin{figure}[t]
\centering
\includegraphics[width=0.45\textwidth]{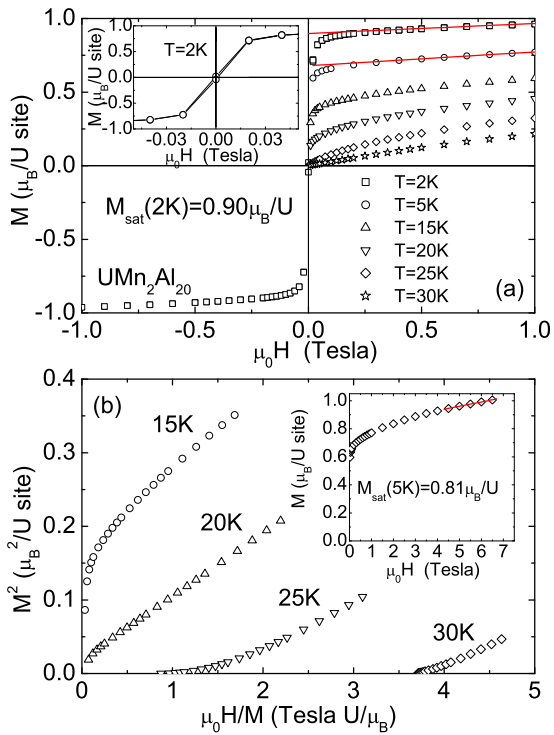}
\caption{\label{fig:2} (a)Magnetization of UMn$_2$Al$_{20}$ at 2 K,
5 K, 15K, 20K, 25K and 30K. The extrapolations of the solid lines
give the saturation magnetization. Inset shows the hysteresis loop
at 2 K. (b) Arrott plots for several temperature. Inset shows the
magnetization at 5 K at larger fields.} \vspace*{-3.5mm}
\end{figure}

The isothermal magnetization results at various temperature of
UMn$_2$Al$_{20}$ are displayed in Fig. 2(a). The full hysteresis
loop at 2 K is shown on a zoomed scale in the inset. Both the
coercive field and the remnant magnetization are very small with
$H_c$ $\sim$ 9 Oe and $M_R$ $\sim$ 0.03 $\mu_B$. A linear fit to the
magnetization data at 2 K below 1 tesla gives a value $M_{sat}$(2K)
= 0.90 $\mu_B$ for the saturation magnetization which is essentially
the same as the value $M_0$ = 0.89 $\mu_B$ derived from the mean
field fit of Fig. 1(b), inset. A similar extrapolation performed on
the 5 K magnetization data below 1 tesla yields $M_{sat}$(5K) = 0.68
$\mu_B$ while the value extrapolated from high field (4.5 T to 6.5
T) is $M_{sat}$(5K) = 0.81 $\mu_B$. We note that these values are
much smaller than the values expected for the $J$ = 4 (5$f^2$, 3.58
$\mu_B$) or $J$ = 9/2 (5$f^3$, 3.62 $\mu_B$) free ions. An Arrott
plot\cite{Arrott} is displayed in Fig. 2(b): it clearly shows that the
Curie temperature $T_c$ is 20-21K.

The specific heat measurements on UMn$_2$Al$_{20}$ and the
nonmagnetic counterpart ThV$_2$Al$_{20}$ are shown in Fig. 3(a).
There is no obvious anomaly in the as-measured data of
UMn$_2$Al$_{20}$ near $T_c$. In the inset, a fit to the form $\gamma
T + \beta T^3$ yields a linear coefficient 0.3 J/mole K$^2$ and a
Debye temperature $\theta_D =$ 337 K.  The magnetic contribution to
the specific heat C$_{mag}$ is obtained by subtracting the lattice
contribution which is equated to the specific heat of the
nonmagnetic counterpart ThV$_2$Al$_{20}$. Both C$_{mag}$ and
C$_{mag}/T$ are shown in Fig. 3(b). The data for  C$_{mag}$ shows a
broad peak near 16 K which corresponds to a small anomaly (a
curvature change) in C$_{mag}/T$ at the same temperature. The
entropy associated with the magnetic specific heat is shown in Fig.
3(c), giving a value for the magnetic entropy of Rln2 at 46 K and
showing a curvature change near $T_c$.

\begin{figure}[t]
\centering
\includegraphics[width=0.5\textwidth]{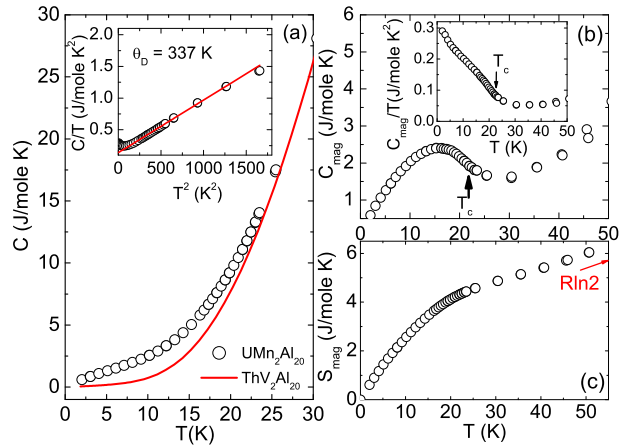}
\caption{\label{fig:3} (a) Specific heat of UMn$_2$Al$_{20}$ and
ThV$_2$Al$_{20}$. The inset shows $C/T$ vs. $T^2$; the solid line is
a linear fit in the temperature range of 15 K to 40 K. (b) Magnetic
contribution to the specific heat. The inset is $C_{mag}/T$ vs. temperature.
(c) Magnetic entropy associated with $C_{mag}$. } \vspace*{-3.5mm}
\end{figure}

The resistivity $\rho (T)$ of UMn$_2$Al$_{20}$ is shown in Fig. 4.
The resistivity decreases with the decreasing temperature down to 10
K, below which it is a constant. There is no anomaly at 20 K
associated with the ferromagnetic transition. We also display the
temperature differential curve d$\rho(T)$/d$T$ in the inset (a) to
enhance the possibility of observing a tiny anomaly in $\rho (T)$.
There is still no obvious anomaly. We combine a
Bloch-Gr$\ddot{u}$neisen resistivity $\rho_{BG}$ together with a parallel
resistor $\rho_P$ to fit $\rho (T)$ as:\\
$\rho (T)^{-1}$=$\rho_P^{-1}+(\rho_0+\rho_{BG})^{-1}$\\
where $\rho_0$ is the residual resistivity. The solid line
represents the best fit to the data. The fit gives $\theta_D$ = 320
K, close to the value $\theta_D$ obtained from the specific heat,
$\rho_0$=61.7 $\mu\Omega$ $cm$, and $\rho_P$=112.8 $\mu\Omega$ $cm$.
This form of resistivity and the magnitude of $\rho_P$ is
characteristic of many transition metal and actinide compounds,
where the parallel resistivity gives rise to a saturation of the
resistivity at a value where the mean free path is comparable to the
lattice spacing. Recent theory\cite{Presistor} indicates that this
saturation happens when the electron phonon interaction destroys
lattice periodicity and momentum conservation at elevated
temperatures. For UMn$_2$Al$_{20}$, the point of the fit is that the
resistivity arises primarily from the electron-phonon interaction,
with little indication of magnetic scattering.

The most interesting property of this compound is that the
magnetization measurements show clearly a ferromagnetic transition
while no obvious anomaly associated with the transition is seen in
the as-measured specific heat or the resistivity. Similar behavior
is observed in the weak itinerant ferromagnetic compound
ZrZn$_2$\cite{ZrZn2}.  The specific heat coefficient $C_{mag}/T$ of
UMn$_2$Al$_{20}$ = 0.3 J/mole K$^2$ is large, suggesting that the
ferromagnetic order occurs within a heavy Fermion state. Together
with the moderately small moment of 0.90 $\mu_B$, the similarity to
ZrZn$_2$ suggests that this system may be a heavy Fermion itinerant
ferromagnet. In this scenario, the reduced entropy and specific heat
anomaly at $T_c$ occurs because the entropy is already small due to
the reduction of the moment by Kondo-like processes.

The Pauli paramagnetism seen for UV$_2$Al$_{20}$ and
UMo$_2$Al$_{20}$ in Fig. 1(c) is also seen in
LnT$_2$Al$_{20}$(Ln=La, Ce and Eu, T=Ti, Mo and V)\cite{LnT2Al20}.
This suggests that a tendency for the $f$ electron to be
non-magnetic is preferred in this structure. This lends further
support to the scenario that the ground state of UMn$_2$Al$_{20}$ is
essentially that of a weakly ferromagnetic itinerant heavy Fermion
compound.

\begin{figure}[t]
\centering
\includegraphics[width=0.45\textwidth]{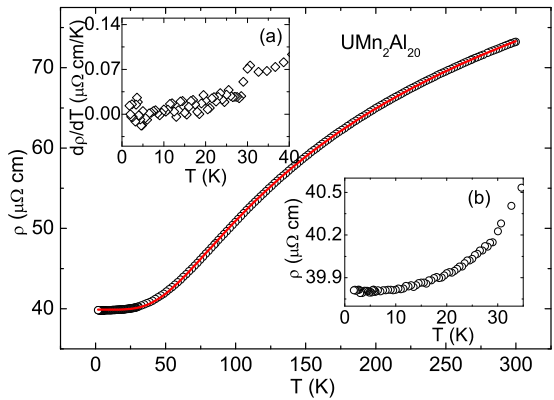}
\caption{\label{fig:4} (a) The resistivity $\rho (T)$ of
UMn$_2$Al$_{20}$. The solid line is a fit to the parallel resistor
model as described in the text. Inset (a) is the temperature
differential curve and (b) is $\rho (T)$ below 35 K; no anomaly is
observed at $T_c$. } \vspace*{-3.5mm}
\end{figure}

The small anomalies in $C(T)$ and $\rho(T)$ were also observed in
Pr$_3$Tl\cite{Pr3Tl} and Pr$_3$In\cite{Pr3In}, where induced
ferromagnetic (antiferromagnetic) order occurs at 12 K\cite{Pr3Tl,
Pr3In}. For these compounds, the Pr$^{3+}$ 4$f^2$ ground multiplet
is split by the crystal field such that the $\Gamma_1$ singlet is
the ground state and the $\Gamma_4$ triplet is the lowest excited
state. The $\Gamma_1$ ground state couples with $\Gamma_4$ triplet
states through the intersite magnetic exchange interaction to induce
a magnetic moment on the ground state\cite{Pr3Tl, Pr3In}. In mean
field theories of the induced magnetic order, the ordering occurs
within the singlet without loss of degeneracy, so that a very weak
anomaly in the specific heat and resistivity is expected, reflecting
the lack of a significant magnetic entropy change at the magnetic
transition temperature. This has been taken as the explanation of
the small anomalies in  $C(T)$ and $\rho(T)$ in Pr$_3$Tl and
Pr$_3$In.

In UMn$_2$Al$_{20}$, the uranium 5$f$ electrons have the possibility
of being in a 5$f^2$ local moment configuration with a nonmagnetic
$J$=0 ground state and a triplet excited state, which is the same
4$f^2$ configuration as in the rare earth Pr$^{3+}$. Coupled with
the absence of a specific heat $C(T)$ and electrical resistivity
$\rho(T)$ anomaly at the transition temperature, this raises the
possibility that this compound has a similar induced local moment
behavior.

It has been proposed that the phase transition in induced
moment systems is brought about by a softening of the crystal field
excitation at the $Q$ vector which corresponds to the magnetically
ordered phase ($Q$ = 0 for ferromagnetism; $Q = Q_N$ for
antiferromagnetism). At a temperature much higher than the ordering
temperature, well-defined non-dispersive crystal field excitations
are expected but in the ordered state the singlet-triplet excitation
would be dispersive\cite{induced}. These effects should be readily
observable in neutron scattering spectra.

\begin{figure}[t]
\centering
\includegraphics[width=0.4\textwidth]{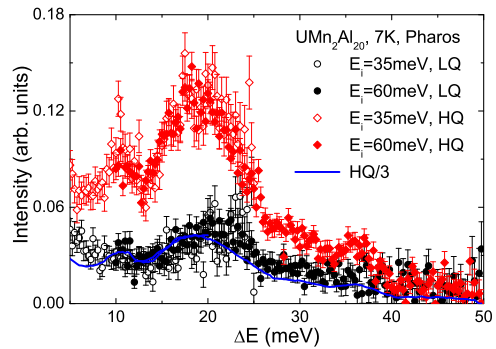}
\caption{\label{fig:5} The comparison of low $Q$ and high $Q$
inelastic neutron scattering spectra for UMn$_2$Al$_{20}$. Error bar
in figure represents $\pm \sigma$ The data were collected on Pharos
at two different incident energies. The solid line is the high $Q$
data divided by a factor of 3.} \vspace*{-3.5mm}
\end{figure}

Therefore, in either the itinerant ferromagnetism case or the local
moment induced ferromagnet case, there should be magnetic
excitations in inelastic neutron scattering spectra corresponding
either to the spin fluctuation (Kondo-like) scattering of the heavy
fermion compound or to the crystal field excitations expected for a
singlet-triplet induced moment system. Unfortunately, in the
inelastic neutron scattering on a polycrystalline sample of
UMn$_2$Al$_{20}$, there are no obvious magnetic excitations and all
the peaks appear to be phonon contributions in the energy range of 5
meV to 50 meV (Fig. 5). Any magnetic excitations at these energies
must overlap the phonon contribution. To estimate the phonon
contribution, we utilize the observation\cite{Bauer1} that in the
neutron scattering spectra for UCo$_2$Zn$_{20}$ and
ThCo$_2$Zn$_{20}$, the phonon contribution at high $Q$ is roughly 3
times larger than at low $Q$. Once the high $Q$ spectra of
UMn$_2$Al$_{20}$ is divided by a factor of 3, it is almost identical
with the low $Q$ spectra, suggesting that if there is magnetic
scattering in this energy range, it is very weak.

In summary, we report a new ferromagnetic compound UMn$_2$Al$_{20}$
for which a clear ferromagnetic transition is observed in the
magnetic susceptibility but no strong anomaly was observed in the
specific heat or resistivity. There appear to be two possible
explanations for this behavior: moderately small moment itinerant
ferromagnetism occurring in a heavy fermion state, or
singlet-triplet induced local moment ferromagnetism. The inelastic
neutron scattering spectra show no obvious magnetic excitations
between 5 meV to 50 meV. More careful neutron scattering experiments
to better determine the nonmagnetic scattering, and to explore the
scattering at lower energies, are in order.

\emph{Acknowledgements} Research at UC Irvine was supported by the
U.S. Department of Energy, Office of Basic Energy Sciences, Division
of Materials Sciences and Engineering under Award DE-FG02-03ER46036.
Work at Los Alamos National Laboratory was performed under the
auspices of the U.S. DOE/Office of Science. Work at ORNL was
sponsored by the Laboratory Directed Research and Development
Program of Oak Ridge National Laboratory, managed by UT-Battelle,
LLC, for the U. S. Department of Energy. We acknowledge the support
of the National Institute of Standards and Technology, U. S.
Department of Commerce, in providing the neutron research facilities
used in this work. Part of the work was supported by the National
Science Foundation under grant no. DMR 0804032.

\end{document}